\documentclass[aps,prl,twocolumn,showpacs,floatfix,superscriptaddress]{revtex4}
\usepackage{graphicx}

\begin{document}
\flushbottom

\title{Evolution of the local superconducting density of states in ErRh$_4$B$_{4}$ close to the ferromagnetic transition}
\author{V. Crespo}
\affiliation{Laboratorio de Bajas Temperaturas, Departamento de
F\'isica de la Materia Condensada, Instituto de Ciencia de
Materiales Nicol\'as Cabrera, Facultad de Ciencias, Universidad
Aut\'onoma de Madrid, 28049 Madrid, Spain}
\author{J.G. Rodrigo}
\affiliation{Laboratorio de Bajas Temperaturas, Departamento de
F\'isica de la Materia Condensada, Instituto de Ciencia de
Materiales Nicol\'as Cabrera, Facultad de Ciencias, Universidad
Aut\'onoma de Madrid, 28049 Madrid, Spain}
\author{H. Suderow}
\affiliation{Laboratorio de Bajas Temperaturas, Departamento de
F\'isica de la Materia Condensada, Instituto de Ciencia de
Materiales Nicol\'as Cabrera, Facultad de Ciencias, Universidad
Aut\'onoma de Madrid, 28049 Madrid, Spain}
\author{S. Vieira}
\affiliation{Laboratorio de Bajas Temperaturas, Departamento de
F\'isica de la Materia Condensada, Instituto de Ciencia de
Materiales Nicol\'as Cabrera, Facultad de Ciencias, Universidad
Aut\'onoma de Madrid, 28049 Madrid, Spain}
\author{D. Hinks}
\affiliation{Materials Science Division, Argonne National Laboratories, Argonne, Illinois 60439}
\author{I.K. Schuller}
\affiliation{Physics Department, University of California-San Diego, La Jolla California 92093-0319, USA}

\begin{abstract}
We present local tunneling spectroscopy experiments in the superconducting and ferromagnetic phases of the reentrant superconductor ErRh$_4$B$_{4}$. The tunneling conductance curves jump from showing normal to superconducting features within a few mK close to the ferromagnetic transition temperature, with a clear hysteretic behavior. Within the ferromagnetic phase, we do not detect any superconducting correlations. Within the superconducting phase we find a peculiar V-shaped density of states at low energies, which is produced by the magnetically modulated phase that coexists with superconductivity just before ferromagnetism sets in.
\end{abstract}

 \date{\today} \maketitle

The physics of competing orders has been the subject of much research over the years. A particularly interesting and extensively studied area is the competition and coexistence between superconductivity and magnetism \cite{BookFischer,ProcGeneva,Bulaevski85,Flouquet02,Buzdin05}. This interest has been reemphasized by advances in highly correlated superconductors, such as the cuprates \cite{Johnston88,Tranquada88}, and, very recently, the Fe pnictide \cite{Kamihara08} and Ni \cite{Watanabe07} and Fe \cite{Kamihara06} phosphide superconductors. In all these materials there is evidence of some sort of magnetism and of superconductivity perhaps at different temperatures or even coexisting. It may even be that for the existence of high temperature superconductivity proximity to a magnetic boundary is important\cite{Orenstein00}. Thus investigating the form in which these two orders interact is not only unusual and interesting, but it may hold information regarding the mechanism of superconductivity in highly correlated systems. A classical example of a superconductor which also exhibits magnetic order is ErRh$_4$B$_4$\cite{Fertig77}. This is a reentrant superconductor in which, with decreasing temperature, first a superconducting transition occurs, at T$_{c1}\approx$ 8 K, and then at T$_{c2}\approx$ 0.7 K, where local Er moments order ferromagnetically, the material becomes normal again. Until now this material was only probed with macroscopic probes, such as thermal studies, resistivity and susceptibility, neutron scattering or tunneling spectroscopy using thing films \cite{Sinha82,BookFischer,ProcGeneva,Bulaevski85,DePuydt86,Umbach81,Poppe81,Prozorov08,Prozorov08b}. The superconducting properties far from the magnetic transition remain unknown to a large extent. For example, the jump in the specific heat at T$_{c1}$ is sizable. However, already shortly below, magnetic contributions are overwhelming. Even more mysterious is the behavior close to ferromagnetism, where always a strong hysteresis is found. When heating, the transition to the full superconducting state occurs T$_{c2\uparrow}$, which is considerably above (around 100 mK) the transition temperature found when cooling, T$_{c2\downarrow}$. There are no pronounced effects associated with the velocity of the temperature ramps, so it seems that superconductivity truly hinders the outcome of ferromagnetism and viceversa \cite{Bulaevski85,Prozorov08,Prozorov08b}. Remarkably, neutron scattering experiments by Sinha et al. \cite{Sinha82} have shown that in the superconducting phase, when cooling, a new magnetically modulated state appears around 1 K, and disappears below T$_{c2\downarrow}$. When heating, the same state appears at T$_{c2\uparrow}$ and disappears again around 1 K \cite{Sinha82,BookFischer}. This peculiar magnetic state is supposed to be induced by superconductivity\cite{Bulaevski85}, as proposed by Anderson and Suhl\cite{Anderson59,Bulaevski85}. More recently, it has been shown that the modulated state most likely coexists within superconducting domains forming close to ferromagnetism\cite{Prozorov08,Prozorov08b}. There are many important open questions raised by these and other studies. Among them, we may ask: What is the shape of the superconducting density of states? What changes are produced in the coexistence region by the magnetically modulated state? Are there some superconducting correlations in the electronic density of states below T$_{c2\downarrow}$ or T$_{c2\uparrow}$? Here we present atomic resolution scanning tunneling microscopy and spectroscopy (STM/S) experiments on a ErRh$_4$B$_4$ single crystal, which provide new insight into these problems.

\begin{figure}[ht]
\includegraphics[width=8cm,clip]{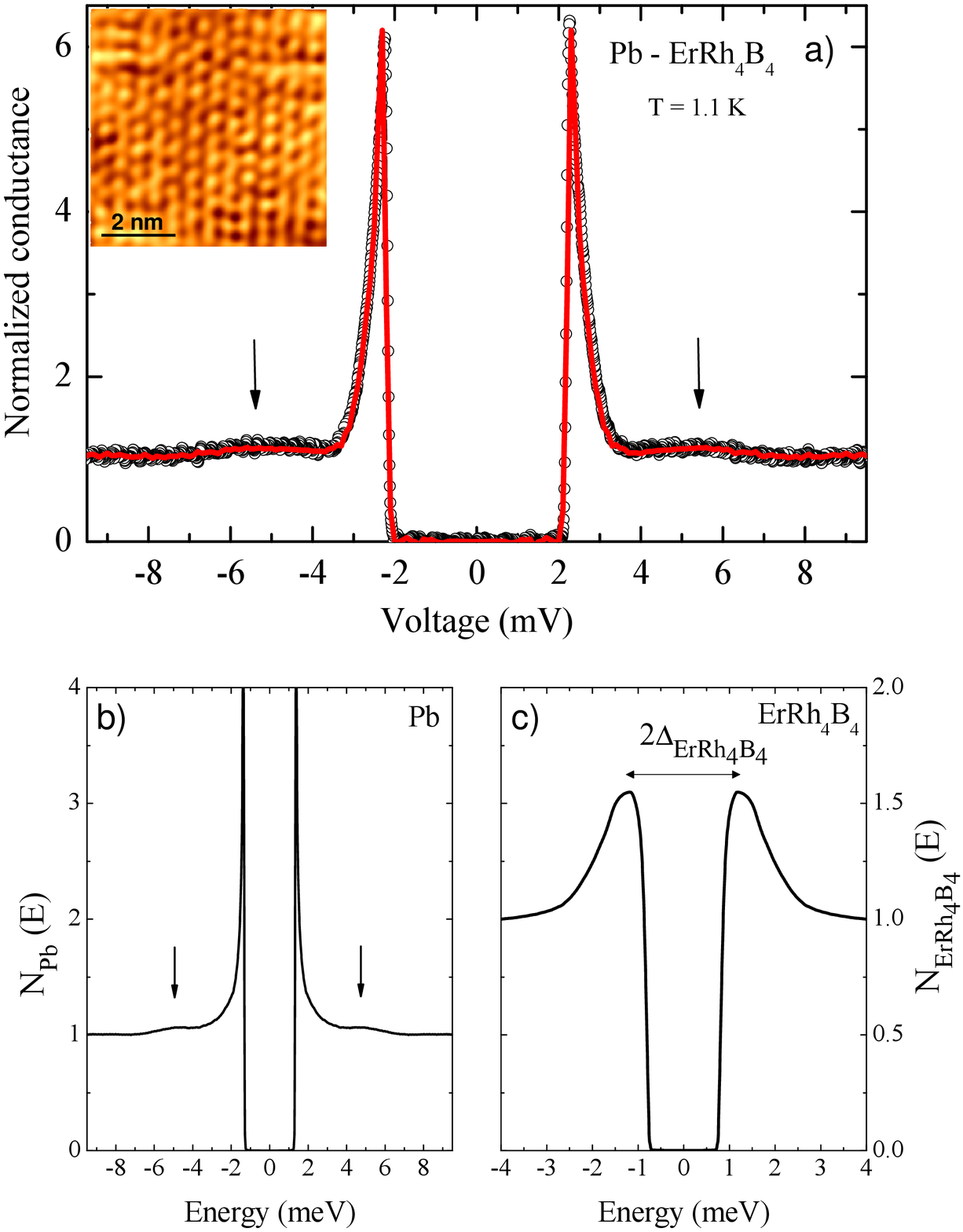}
\caption{(Color online) In (a) we show local tunneling spectroscopy curves obtained in the superconducting phase of ErRh$_4$B$_4$ at 1.1 K with a Pb tip, and in the inset atomic resolution topography. Red line is the calculated tunneling conductance using $N_{Pb}(E)$ and $N_{ErRh_4B_4}(E)$ shown, respectively, in (b) and (c). Arrows in (a) and (b) mark Pb phonon mode features.}\label{Fig1}
\end{figure}

We use a STM/S system in a $^3$He refrigerator, equipped with a Pb tip and an in-situ system, described previously, to obtain clean and sharp tips\cite{Rodrigo04,Rodrigo04c}. The superconducting density of states of Pb tips has been discussed in previous work\cite{Suderow02,Rodrigo04,Rodrigo04c}. As it is well known, tunneling spectroscopy on a superconducting sample using a superconducting counter electrode (S-S' tunneling) is superior to tunneling spectroscopy with a normal electrode, because the sharp density of states of the counter electrode allows for a better determination of fine structure in the density of states of the sample, even at temperatures where thermal smearing is important. The tunneling current between a superconducting tip with density of states $N_{Pb}(E)$ and a sample with $N_{ErRh_4B_4}(E)$ can be written as $I(V) \propto \int dE [f(E-eV)-f(E)] N_{Pb}(E-eV) N_{ErRh_4B_4}(E)$, where $f(E)$ is the Fermi function. At all temperatures $I(V)$, and its derivative the tunneling conductance $\sigma(V)$, present sharp features at $|V|=\Delta_{Pb}+\Delta_{ErRh_4B_4}$, and at temperatures above approximately $T_c/3$ at $\Delta_{Pb}-\Delta_{ErRh_4B_4}$\cite{Rodrigo04,Rodrigo04c,Guillamon07}. With the accurately previously determined density of states of the Pb tip as a function of temperature, $N_{Pb}(T;E)$, it is possible to obtain the density of states of the sample $N_{ErRh_4B_4}(T;E)$ from the previous expression\cite{Guillamon07}.

The sample is a single crystal in the primitive tetragonal phase grown from an Er-Rh-B melt. The ingot contained two large single crystals ($\approx$0.3 g), the larger of which was used for the present studies and is the same earlier used \cite{Sinha82} for combined neutron, transport and magnetic studies. We measured the single crystal on differently oriented surfaces, in and out of plane of the primitive tetragonal crystal structure. In all cases, tunneling characteristics found on as-grown surfaces, correspond to good vacuum tunnel junction, with reproducible topographic images and spectra, independent on the bias voltage, and work functions of the order of some eV. The topography of the sample is irregular, and it appears difficult to find extended flat regions. Many regions show depressed superconducting properties, and even no superconductivity at all, possibly due to changes in the composition of the surface, surface reconstructions, or strongly enhanced magnetic scattering. However, we were able to find some (around twenty) locations with flat areas of some 100 nm x 100 nm, where we often obtain atomic resolution images (inset of Fig.\ref{Fig1}). These areas show the spectroscopic features discussed in the following. These features are much sharper than those found on macroscopic measurements\cite{Sinha82,BookFischer,ProcGeneva,Bulaevski85,DePuydt86,Umbach81,Poppe81,Prozorov08,Prozorov08b}, which should, even in high quality samples, give an averaged behavior.

\begin{figure}[ht]
\includegraphics[width=7cm,clip]{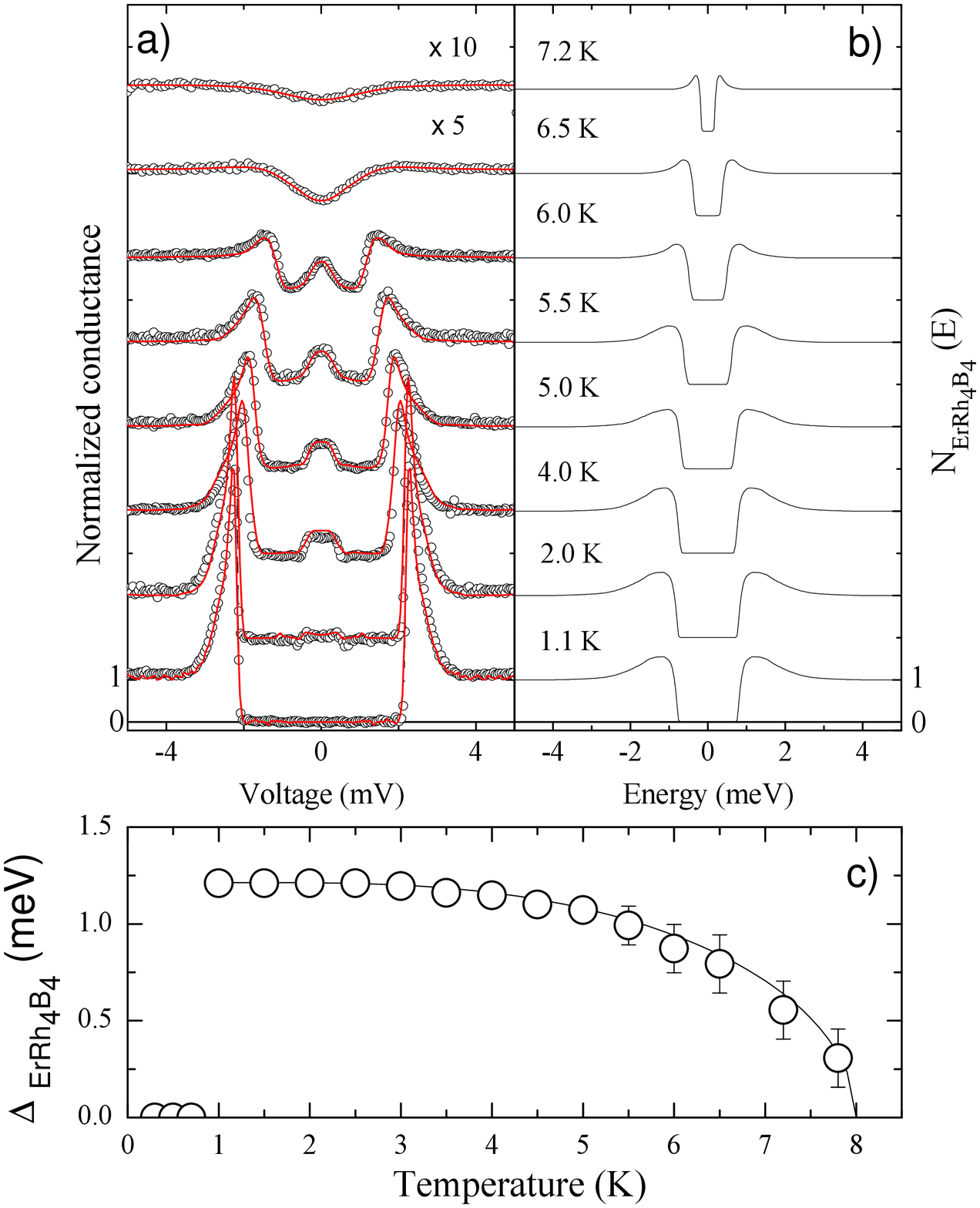}
\caption{(Color online) In (a) we show the temperature dependence of the tunneling conductance curves measured (points) and those calculated (lines) using $N_{ErRh_4B_4}(E;T)$ shown in (b) (curves are shifted for clarity; scale in the top two curves is enlarged). In (c) we show the temperature dependence of the voltage position of the maximum in $N_{ErRh_4B_4}(E)$, $\Delta_{ErRh_4B_4}$, together with the BCS curve (line).}
\label{Fig2}
\end{figure}

At the lowest temperatures, T=0.3 K, only the superconducting features due to the Pb tip are observed in the tunneling conductance curves, and $N_{ErRh_4B_4}(E)$ is flat and featureless. Well within the superconducting phase of ErRh$_4$B$_4$ we measure the expected curves characteristic for tunneling between two superconductors [Fig.\ref{Fig1}(a)]. Using N$_{Pb}(E)$ (Fig.\ref{Fig1}(b) and Ref.\cite{Rodrigo04b}), we obtain the density of states of ErRh$_4$B$_4$, $N_{ErRh_4B_4}(E)$,  shown in Fig.\ref{Fig1}(c). $N_{ErRh_4B_4}(E)$ has a well opened superconducting gap, with a zero density of states up to 0.75 meV, and a rounded quasiparticle peak whose maximum is located at $\Delta_{ErRh_4B_4}$ =1.2 meV. As shown in Fig.\ref{Fig2}, when heating, the local tunneling conductance curves show S-S' features, with peaks appearing at $\Delta_{Pb}-\Delta_{ErRh_4B_4}$ from thermal excitations. The rounded shape of $N_{ErRh_4B_4}(E)$ is maintained as a function of temperature. $\Delta_{ErRh_4B_4}$ presents a temperature evolution (Fig.\ref{Fig2}) which is very close to the expected behavior from weak coupling BCS theory ($\Delta_{BCS}$ = 1.76 k$_B$T$_{c1}$ = 1.2 meV).

The rounded quasiparticle peaks in $N_{ErRh_4B_4}(E)$ are very different from the divergency expected within single band s-wave BCS theory. Most probably, this shows peculiar, temperature independent, magnetic pair breaking effects from disorder in the Er paramagnetic sublattice \cite{Brison04,Crespo06a,Gusakova06,Coffey83}. However, we cannot exclude an intricate dependence of the superconducting gap over the Fermi surface. Band structure calculations show that the largest contribution to the Fermi level density of states comes from the 4d electrons, with smaller additional p and s character contributions, of the Rh atoms, and a contribution from Er 5d electrons\cite{Jarlborg77}. In any case, the close agreement between experiment and BCS theory shown in Fig.\ref{Fig2}(c), and the fact that the Pb phonon modes are displaced in the tunneling conductance curves by exactly $\Delta_{ErRh_4B_4}/e$ (Fig.\ref{Fig1}), pinpoint that the main opened gap feature over the Fermi surface is that of $\Delta_{ErRh_4B_4}$.

\begin{figure}[ht]
\includegraphics[width=7cm,clip]{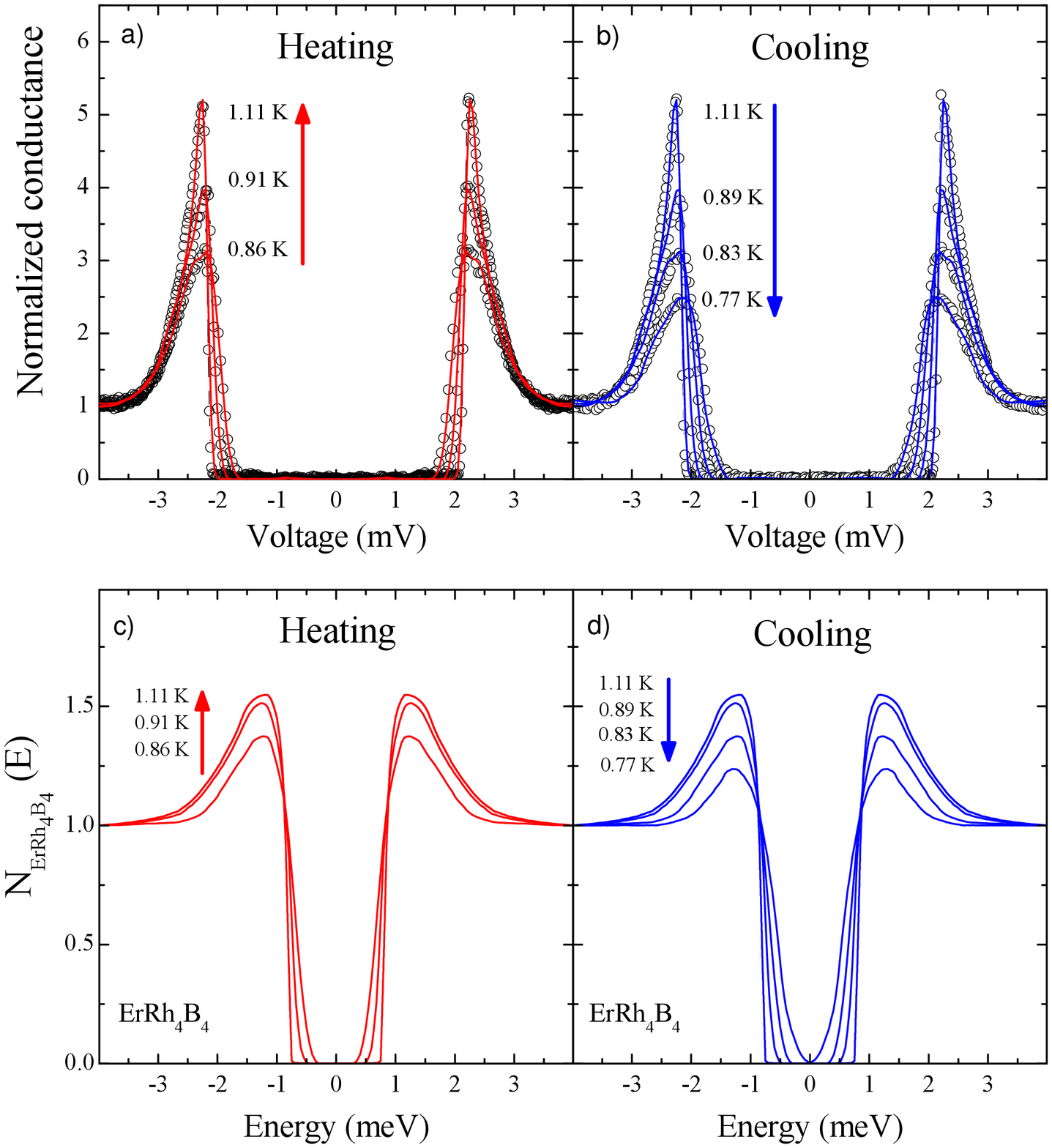}
\caption{(Color online). A close-up view of experiments made near the ferromagnetic transition. In the left panels, behavior when heating, from 0.86 K to 1.1 K, and in right panels, when cooling from 1.1 K to 0.77 K. In (a) and (b) we show the tunneling spectroscopy curves obtained, together with those calculated using $N_{ErRh_4B_4}(E)$ shown in (c) and (d).}
\label{Fig3}
\end{figure}

\begin{figure}[ht]
\includegraphics[angle=270,width=8cm,clip]{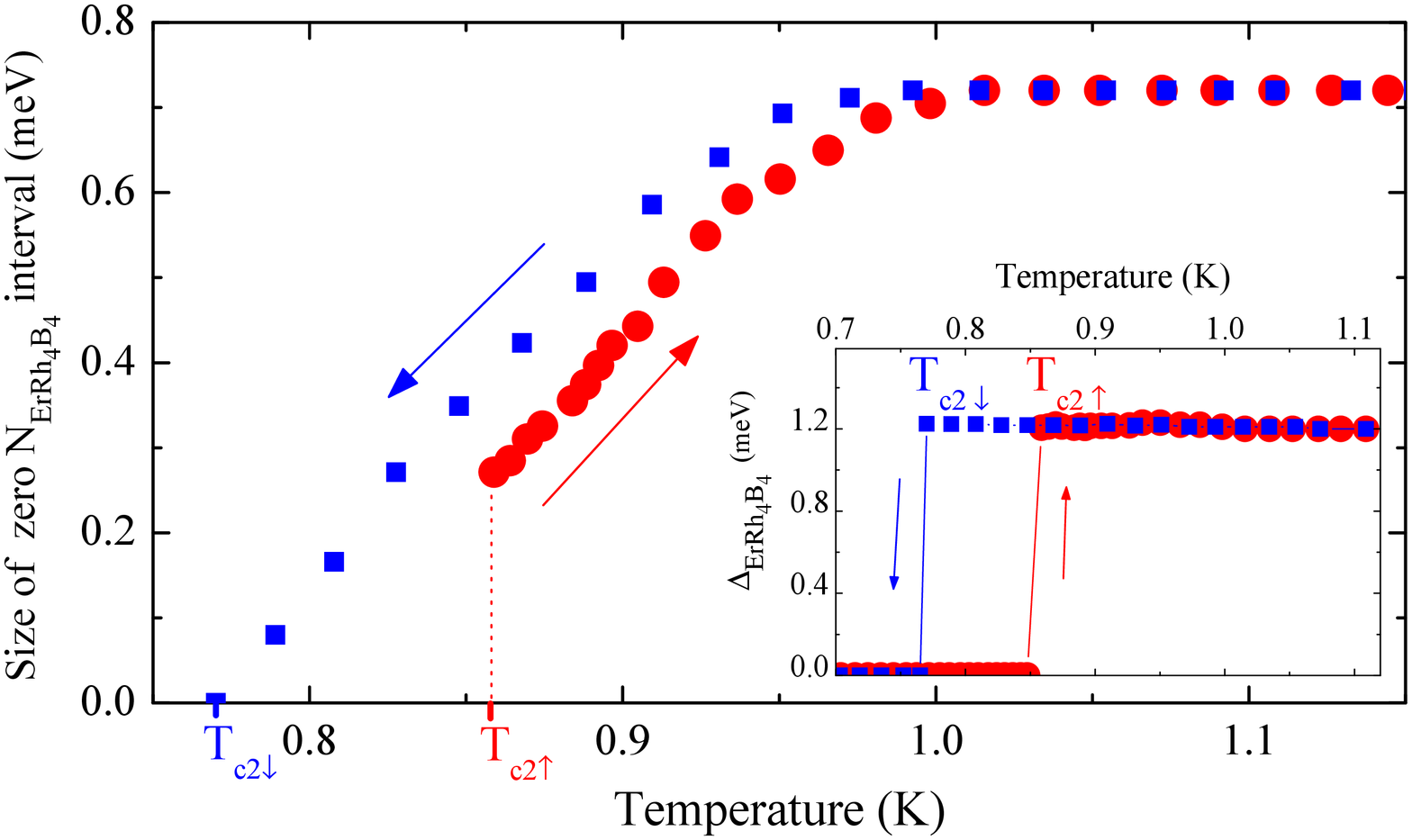}
\caption{(Color online). The width of the zero density of states interval is shown as a function of temperature when cooling and heating (respectively, blue squares and red circles). The inset shows $\Delta_{ErRh_4B_4}(T)$ close to the magnetic transition.}
\label{Fig4}
\end{figure}

Most remarkable features in $N_{ErRh_4B_4}(E)$ are found close to the ferromagnetic transition, where the tunneling conductance curves strongly change its shape. In Fig.\ref{Fig3} we show a close-up view of typical heating and cooling experiments. When heating, superconductivity in the sample appears abruptly at T$_{c2\uparrow}$, and the tunneling spectroscopy curves show, within a few mK, the S-S' behavior represented by the lowest curve in Fig.\ref{Fig3}(a). The resulting temperature dependence of $N_{ErRh_4B_4}(E)$ is shown in Fig.\ref{Fig3}(c). Close to T$_{c2\uparrow}$, at T = 0.86 K in Fig.\ref{Fig3} (a) and (c), the curves are significantly different than well within the superconducting phase, at T = 1.1 K in Fig.\ref{Fig3} (a) and (c). The position of the maximum of the quasiparticle peaks is still at $\Delta_{ErRh_4B_4}$. However, the quasiparticle peaks are smeared and the energy interval with zero $N_{ErRh_4B_4}(E)$ is of about 0.3 meV, i.e. smaller than well within the superconducting phase (0.75 meV, see Figs.\ref{Fig1} and \ref{Fig4}).

When cooling again into the ferromagnetic phase, we find identical tunneling conductance curves, at, however, lower temperatures than in the heating process, as shown in Fig.\ref{Fig3}(b), and the disappearance of superconducting features when cooling occurs at T$_{c2\downarrow}$, which is 90 mK below T$_{c2\uparrow}$. The transition to the normal state occurs again abruptly, within a few mK. Just before the transition, at a few mK above T$_{c2\downarrow}$, we find tunneling conductance curves with most strongly smeared superconducting features [see curves at 0.77 K in Figs.\ref{Fig3}(b) and (d)]. Remarkably, the interval with zero $N_{ErRh_4B_4}(E)$ is fully lost close to T$_{c2\downarrow}$, where we get truly gapless superconductivity with V shaped increase of $N_{ErRh_4B_4}(E)$ at low energies. Such a behavior is never observed close to T$_{c2\uparrow}$ when heating, as highlighted in Fig.\ref{Fig4}, where we plot the temperature dependence of the energy interval with zero $N_{ErRh_4B_4}(E)$. Note that this interval is significantly influenced by magnetism up to about 0.2 K above the appearance of the ferromagnetic state, whereas $\Delta_{ErRh_4B_4}$ (inset of Fig.\ref{Fig4}) remains constant.

These results have been reproduced in the locations where we focus on here, making measurements at constant temperature with steps as small as 2 mK. At a given temperature and location, the tunneling conductance curves are spatially homogeneous, and do not change as a function of time. There are small differences in the transition temperatures and height and position of the quasiparticle peaks in different locations, possibly due to internal stress. Nevertheless, the difference in transition temperatures, T$_{c2\uparrow}$-T$_{c2\downarrow}$, the abrupt nature of the transition, the fact that the position of the quasiparticle peaks remains at $\Delta_{ErRh_4B_4}$, and the gapless V shaped increase of $N_{ErRh_4B_4}(E)$ close to T$_{c2}$ are always found.

The possible coexistence between long range ferromagnetic order and superconductivity has been discussed by comparing different macroscopic experiments\cite{BookFischer,ProcGeneva,Bulaevski85}. As shown in Refs.\cite{Sinha82}, the height of the ferromagnetic Bragg peaks jumps at temperatures close to T$_{c2\uparrow}$ and T$_{c2\downarrow}$, showing the same hysteretic behavior as we find here. Moreover, it does not saturate but has a continuous increase below these temperatures\cite{Sinha82}. The full disappearance of any \emph{local} superconducting signal in the tunneling spectroscopy data at T$_{c2\downarrow}$, and the corresponding appearance at T$_{c2\uparrow}$, shows that there is no evidence for coexistence between long range ferromagnetic order and superconductivity. If there would be superconducting correlations in extended ferromagnetic regions, these should lead to some signal in the local tunneling conductance curves below T$_{c2\uparrow}$ and T$_{c2\downarrow}$, such as a decrease of $N_{ErRh_4B_4}(E)$ close to the Fermi level\cite{Bulaevski85}, which we do not observe here.

The presence of the magnetically modulated state discovered in Ref.\cite{Sinha82} best explains the observed smearing in the tunneling density of states. The peaks satellite to some of the main ferromagnetic Bragg reflections, which correspond to a modulated magnetic moment at (0.042a,0.055c), exist in the superconducting phase on the same temperature range where we observe the changes in the width of the zero $N_{ErRh_4B_4}(E)$ interval (Fig.\ref{Fig4}). Moreover, the satellite Bragg peaks grow to much higher values on cooling at T$_{c2\downarrow}$ than on heating at T$_{c2\uparrow}$, and disappear abruptly when entering long range ferromagnetic order\cite{Sinha82}. Clearly, the smearing of the density of states found here (Fig.\ref{Fig4}) must directly show the effect of this kind of magnetic order on the superconducting density of states. This is strongest when the moments associated with the magnetically modulated state are highest (at T$_{c2\downarrow}$) and where we observe the peculiar gapless regime.

The superconducting density of states in the magnetically modulated phase has been qualitatively calculated previously\cite{Bulaevski83,Bulaevski85}. The predictions coincide with our observations. Within this scenario, while the main gap parameter of the superconducting phase $\Delta_{ErRh_4B_4}$ remains unchanged, the presence of a finite magnetization in some directions leads to selective pair breaking effects which produce strongly anisotropic gap structures similar to those observed in d-wave superconductors. In particular, there are zero gap regions along the lines formed by equally oriented magnetic moments, although no sign changes of the phase of the Cooper pair wavefunction, and no concomitant zero energy bound states (characteristic of d-wave superconductivity, see e.g. Ref.\cite{Pan00}), are found.

In summary, local tunneling spectroscopy experiments have allowed us to find locations in the surface of ErRh$_4$B$_4$ with areas showing clear-cut superconducting features in the density of states and a superconducting gap parameter close to BCS expectations. The temperature evolution of the superconducting density of states follows the temperature dependence of magnetic signals found in previous neutron scattering experiments. Ferromagnetism seems to totally cancel superconductivity, and the magnetically modulated phase has a strong effect on the superconducting density of states, which leads to fully gapless superconductivity.

We acknowledge conversations with A.I. Buzdin, F. Guinea, H. Suhl and S.K. Sinha. The Laboratorio de Bajas Temperaturas is associated to the ICMM of the CSIC. This work was supported by the Spanish MICINN (Consolider Ingenio Molecular Nanoscience CSD2007-00010 program and FIS2008-00454), by the Comunidad de Madrid through program "Science and Technology at Millikelvin", by NES and ECOM programs of the ESF, and by the US Department of Energy.


\end{document}